\documentclass[
preprint,
showpacs,
showkeys
]{revtex4-1}
\usepackage{amsthm}
\theoremstyle{plain}
\newtheorem*{theorem*}{Theorem}

\newtheorem*{definition*}{Definition}

\newtheorem*{lemma*}{Lemma}

\usepackage{braket}
\usepackage{delarray}
\usepackage{here}

\usepackage{txfonts}

\usepackage[dvipdfmx]{graphicx}
\usepackage{bm}
\usepackage{color}
\usepackage{ulem}

\newcommand{\be}{\begin{eqnarray}}
\newcommand{\ee}{\end{eqnarray}}
\newcommand{\ba}{\begin{array}}
\newcommand{\ea}{\end{array}}
\newcommand{\bmat}{\left(\begin{array}}
\newcommand{\emat}{\end{array}\right)}

\begin{document}
\title{
Time evolution of the von Neumann entropy in open quantum systems}

\author{Kohei Kobayashi$^1$}

\affiliation{
$^1$Global Research Center for  Quantum Information Science, National Institute of Informatics,
 2-1-2 Hitotsubashi,  Chiyoda-ku, Tokyo 101-8340, Japan}

\begin{abstract} 

Control of open quantum dynamics is of great interest for realizing quantum technologies.
Therefore, it is an important task to quantify and characterize the entropy for open quantum systems under decoherence.
In this paper, we study the time evolution of the von Neumann entropy for open quantum systems described by the Lindblad master equation.
Note that, in particular, when the decoherence corresponds to the measurement for the observable in the system,
the von Neumann entropy tends to monotonically increases as the variance becomes larger.
Furthermore, we present a lower bound of the von Neumann entropy in the long-time limit.
This lower bound has advantages of being straightforwardly calculated and applicable to a general Markovian open quantum system.

\keywords{von Neumann entropy, open quantum system, Lindblad master equation}

\end{abstract}
\date{\today}
\maketitle

\section{Introduction}

Control of quantum dynamics is of significant interest for realizing quantum technologies.
Because real quantum systems are subjected to the interaction with uncontrollable environments, 
 an open quantum system offers a powerful tool for modeling actual phenomena in quantum mechanics \cite{Nielsen}.
Due to the interaction between system and environments, the loss of energy and phase informationm, 
so-called decoherence, is inevitably caused.
Therefore, it is important to characterize how quantum information vary in an open quantum system.

The calculation of entropy in open quantum systems is generally an essential task.
Obtaining a limit on the rate of entropy change or quantifyig the entropy itself enable us to understand the dynamical behavior
 of nonequilibrium quantum states, as a result, which leads to ideal control toward desired states.

Towards this problem, the von Neumann entropy in open quantum systems has already been discussed almost 30 years ago \cite{Benatti}.
In recent years, several results for studying the entropy in open quantum dynamics have been reported 
\cite{Audenaert, Ou, Abe, Bakh, Mohan};
for instance, the time derivative of the von Neumann entropy in a time-dependent harmonic oscillator was studied \cite{Ou},
the lower bound on the Re'nyi entropy rate was given \cite{Abe}, and 
the relation of the von Neumann entropy and the quantum speed limit was investigated \cite{Mohan}.
However, these results remain as a formal mathematical expression 
and physical intuition is hard to extract, and also, 
there has been no result giving any estimate on the entropy itself.

Motivated by the above background, we study the time evolution of the 
von Neumann entropy for open quantum systems descrived by the Lindblad master equation.
From an obtained inequality for the time derivative of the von Neumann entropy, 
we derive a condition that the entropy in the quantum system monotonically increases under decoherence.
Note that, in particular, when the decoherence is the measurement for the observable of the system,
this condition is characterized by the variance of it.
Moreover, we present a lower bound of the von Neumann entropy of an
open quantum system in the long-time limit.
This bound gives a limit on how low the entropy in the quantum system 
can maintain in the presence of decoherence. 
In addition, as a mathematical tool, 
this bound has advantages of being easy to calculate and applicable to a general Markovian open quantum system.

\section{Entropy bound}

\subsection{Controlled quantum dynamics}

Let $\mathcal{H}$ denote the Hilbert space, where the quantum state is defined.
 The state of a quantum system is represented by the density operator $\rho$, 
which satisfies following properties: $\rho=\rho^\dagger$, $\rho\geq 0$, and ${\rm Tr}(\rho)=1$.
The identity operator is denoted by $I$.
Let us assume that the qauntum state $\rho$ is Markov 
(i.e., the system is weakly coupled to the environment and the dynamics of the environment changes much faster than the system).
Therefore, the information only flows from the system to the environment and there is no information flowing back.
In this case, the time evolution of the Markovian open quantum system $\rho_t$ is described by
 the following equation \cite{Lindblad}:

\begin{equation}
\label{me}
\frac{d\rho_t}{dt} = -i[ H , \rho_t] + \mathcal{D}[L ]\rho_t, 
\end{equation}
which is referred as the Lindblad master equation.
 $H$ is the system Hamiltonian, and
 $L$ is the Lindblad operator representing the decoherence process, and thus 
$\mathcal{D}[A]\rho=A\rho A^\dagger-A^\dagger A\rho/2-   \rho A^\dagger A/2$ (we set $\hbar=1$ in the following).
The Lindblad master equation is known as the most general generator of the quantum dynamical semigroup \cite{Heinz}.

\subsection{Main result}

We consider the von Neumann entropy $S_t$:

\begin{equation}
\label{vne}
S_t=-{\rm Tr}\left( \rho_t \ln\rho_t\right), \ \ \ 0\leq S_t \leq \ln d,
\end{equation}
where $d$ is the rank of the quantum system.
The von Neumann entropy quantifies the amount of information contained 
in a state when many identical and independent copies of the state are available.
$S_t$ takes zero when $\rho_t$ is pure, and the maximum is achieved only when the system is maximally mixed $\rho_t=I/d$.  
Because $\rho_t$ can be diagonalized by a certain orthonormal basis $\{ |u_j \rangle\}_j$ as
$\rho_t=\sum_j\lambda_j |u_j\rangle\langle u_j|$, the von Neumann entropy is also written as
\begin{equation}
S_t=-\sum_j \lambda_j \ln \lambda_j,
\end{equation}
which is analogous to the Shannon entropy in classical information theory \cite{Shannon}.

The time evolution of the von Neumann entropy is calculated as follows:

\begin{eqnarray}
\label{dS/dt}
\frac{ dS_t}{dt}   
&=&  - {\rm Tr}\left( \frac{ d\rho_t}{dt} \ln\rho_t\right)    \nonumber  \\
&=&- {\rm Tr}\left\{ (-i[H, \rho_t] +\mathcal{D}[L]\rho_t )   \ln\rho_t \right\}   \nonumber  \\
&=&- {\rm Tr}\left( L \rho_t L^\dagger  \ln\rho_t \right)
+\frac{1}{2}  {\rm Tr}\left( L^\dagger L \rho_t \ln\rho_t \right)
+\frac{1}{2}  {\rm Tr}\left( \rho_t L^\dagger L  \ln\rho_t \right)  \nonumber  \\
&=&   {\rm Tr}\left(L^\dagger L \rho_t \ln\rho_t  \right)
-   {\rm Tr}\left(  L \rho_t L^\dagger \ln\rho_t  \right),
\end{eqnarray}
where we used ${\rm Tr}(d\rho_t/dt)=0$, $[\rho_t, \ln\rho_t]=0$ and  ${\rm Tr}\left( [H, \rho_t]\ln \rho_t \right)=0$.

Here, due to the fact that $-\rho_t\ln\rho_t$ is positive semidefinite, by using the inequality
${\rm Tr}(AB)\leq {\rm Tr}(A){\rm Tr}(B)$ for $A$, $B\geq0$,
 the first term of Eq. (\ref{dS/dt}) can be bounded as follows:

\begin{eqnarray}
\label{first}
  {\rm Tr}\left(  L^\dagger L \rho_t\ln\rho_t  \right) 
&=&  - {\rm Tr}\left[ L^\dagger L (-\rho_t\ln\rho_t)  \right]  \nonumber  \\
&\geq& -{\rm Tr}(L^\dagger L)S_t =\| L\|^2_{\rm F}S_t, 
\end{eqnarray}
where $\| A\|^2_{\rm F}:= {\rm Tr}(A^\dagger A)^{1/2}$ is the Frobenius norm.
Next, to calculate the lower bound of Eq. (\ref{dS/dt}),  we introduce the following inequality:
\begin{equation}
-\ln\rho\geq I-\rho.
\end{equation}
We give the proof of this inequality in Appendix A.
Using it, the second term of Eq. (\ref{dS/dt}) can be bounded as follows:

\begin{eqnarray}
  {\rm Tr}\left[  L \rho_t L ^\dagger (-\ln\rho_t)  \right]
 &\geq & {\rm Tr}\left[  L \rho_t L ^\dagger (I -\rho_t )  \right]  \nonumber  \\
 &\geq &  {\rm Tr}\left(L \rho_t L^\dagger\right) - {\rm Tr}\left(L \rho_t L^\dagger \rho_t \right) \geq 0.
\end{eqnarray}

Therefore, we have the inequality:

\begin{equation}
\label{inequality}
\frac{dS_t  }{dt} \geq - \| L\|^2_{\rm F} S_t 
+  {\rm Tr}\left(L^\dagger L \rho_t \right) - {\rm Tr}\left( L \rho_t L^\dagger \rho_t \right).
\end{equation}
Therefore, it is guaranteed that the time evolution of $S_t$ is always positive when
\begin{equation}
S_t \leq \frac{  {\rm Tr}\left(L^\dagger L \rho_t \right) - {\rm Tr}\left( L \rho_t L^\dagger \rho_t \right) }{ \| L\|^2_{\rm F}  }.
\end{equation}

This upper bound is less than one, because
\begin{equation}
\frac{  {\rm Tr}\left(L^\dagger L \rho_t \right) - {\rm Tr}\left( L \rho_t L^\dagger \rho_t \right) }{ \| L\|^2_{\rm F}  }
\leq \frac{{\rm Tr}\left(L^\dagger L\right) {\rm Tr}\left( \rho_t \right) }{\| L\|^2_{\rm F}  }\leq 1.
\end{equation}
Because $S_t\leq \ln d$, the inequality (9) gives a meaningful result for $d\geq3$.

Furthermore, let us assume that $L$ is self-adjoint  (i.e., $L=L^\dagger$);
 $L$ represents the measurement for an observable in the system. 
In this case,  the numerator of Eq. (9)  is bounded from below as follows:
\begin{eqnarray}
  {\rm Tr}\left(L^\dagger L \rho_t \right) - {\rm Tr}\left( L \rho_t L^\dagger \rho_t \right) 
&=& {\rm Tr}\left( L^2 \rho_t  \right) - {\rm Tr}\left[ (\rho^{1/2}_t L \rho^{1/2}_t )^2 \right]  \nonumber  \\
&\geq&  {\rm Tr}\left( L^2 \rho_t \right) - {\rm Tr}( L\rho_t)^2, 
\end{eqnarray}
where ${\rm Tr}(A^2)\leq {\rm Tr}(A)^2$ for $A\geq 0$ was used.
It is identical to the variance of the observable $L$ in $\rho_t$.
Therefore, denoting ${\rm Var}[L]:={\rm Tr}(L^2\rho_t)- {\rm Tr}(L\rho_t)^2$,
Eq. (9) is rewritten as
\begin{equation}
S_t \leq \frac{ {\rm Var}[L] }{ \| L\|^2_{\rm F}  }.
\end{equation}
This result is reasonable because
the more the measurement outcomes fluctuates, the more easier the conditions for
the von Neumann entropy monotonically increasing.
On the other hand, by suppressing the observable variance small, 
it may be possible to prevent the entropy from monotonically increasing.

Next, 
we show that a fundamental lower bound of $S_t$ can be derived in the limit $t\to\infty$ from the inequality (\ref{inequality}).
To obtain it,
let us consider the function:
\begin{equation}
f(x)=-\| L\|^2_{\rm F} x + {\rm Tr}\left(L^\dagger L \rho_t \right) - {\rm Tr}\left(L \rho_t L^\dagger \rho_t \right).
\end{equation}

$f(0)\geq 0$, and when $x=\ln d$ ( $\rho_t=I/d$), then
\begin{equation}
f(\ln d)= \| L\|^2_{\rm F} \left(\frac{1}{d}- \frac{1}{d^2}-\ln d \right) <0.
\end{equation}

Then, from the above properties of $f(x)$, $f(S_*)=0$ has a unique solution $S_*$ in $[0, \ln d]$.
If $S_T<S_*$ at a given evolution time $T$, $S$ is monotonically increasing.
On the other hand, for the range such that $S_T\geq S_*$, the inequality (7) does not
 say anything about the time evolution
of $S_T$ for $t\geq T$. As a result, in the long-time limit $t\to\infty$, we obtain 
the lower bound:

\begin{equation}
\label{theorem}
S_\infty \geq 
S_*:=\frac{ \beta }{ \| L \|^2_{\rm F} },
\end{equation}
where $\beta:=  
{\rm Tr}\left( L^\dagger L \rho_\infty  \right) - {\rm Tr}\left( L \rho_\infty L^\dagger \rho_\infty \right)$.
That is, $S_*$ gives a limit on how low the entropy in quantum system can maintain under decoherence. 

Here we would like to emphasize the performance of $S_*$ as a tool for quantifying the entropy.
This lower bound is applicable to a general Markovian open quantum system 
without imposing any assumptions on the decoherence $L$.
Also it can be directly calculated once $\rho_\infty$ and $L$ are specified,
thanks to which, there is no need for solving any equation or doing integral calculation.
When the steady state is maximally mixed $\rho_\infty=I/d$, 
$S_*$ is given as the function of $d$ as 
\begin{eqnarray}
S_*\left(I/d \right) = \frac{d-1}{d^2}.
\end{eqnarray}
Then, the lower bound takes the maximum when $d=2$, $S_*=1/4$ and $S_\infty=\ln 2\approx 0.69$,
meanwhile as $d$ increases, its tightness becomes weak.

Finally, the bound (\ref{theorem}) can be generalized to the case where
 the system is sujected to multiple decoherence channels $\sum_j \mathcal{D}[L_j]\rho_t$ as follows:
\begin{eqnarray}
 S_* 
 = \frac{ \sum_j \beta_j }{  \sum_j \| L_j \|^2_{\rm F} }.
\end{eqnarray}

\section{Conclusion}
In this paper, we have derived an inequality for the time derivative of the von Neumann entropy 
in an open quantum system under decoherence.
Based on it, we have characterized the condition that the von Neumann entropy is monotonically increasing.
In particular, it is noted that 
when the decoherence corresponds to the measurement for an observable of the system,
this condition is upper bounded by its variance.
Moreover, we derived a lower bound on the von Neumann entropy itself in the long-time limit.
This lower bound is applicable to a general Markovian open system described by the Lindblad master equation and
can be straightforwardly computed.
The results obtained in this paper are expected to be useful for studying the dynamics of quantum properties in the presence of decoherence.
A remaining work is to further generalize the present discussion to an important dynamics, such as
 non-Markovian dynamics or continuously monitored dynamics \cite{Handel, Geremia}.

This work was supported by MEXT Quantum Leap Flagship Program Grant JPMXS0120351339.

\appendix

\section{Proof of Eq. (6)}

To show the inequality Eq. (6), we use the diagonalization of the density matrix
\begin{equation}
\rho=U\Lambda U^\dagger,
\end{equation}
where $U$ is an appropri
ate unitary matrix and $\Lambda={\rm diag}\{\lambda_1,\cdots,\lambda_d\}$ with the eigenvalues
$1\geq \lambda_1\geq \cdots \geq \lambda_d\geq 0$.
Then,

\begin{eqnarray}
-\ln\rho &=& -U \ln \lambda U^\dagger  \nonumber \\
&=& U {\rm diag}\{ -\ln\lambda_1,\cdots,  -\ln \lambda_d \} U^\dagger  \nonumber \\
&\geq & U {\rm diag}\{ 1-\lambda_1,\cdots,  1-\lambda_d \} U^\dagger  \nonumber \\
&=& U\left(I-\Lambda \right)U^\dagger=I-\rho,
\end{eqnarray}
where we used $-\ln x \geq 1-x$ for $x>0$.


\end{document}